\documentclass{aa}
\usepackage{graphicx}
\usepackage{txfonts}
\begin{document} 

  \title{Beyond small-scale transients: a closer look at the diffuse quiet solar corona}

\author{J. Gorman\inst{\ref{i:mps}}
\and L.P. Chitta\inst{\ref{i:mps}}
\and H. Peter\inst{\ref{i:mps}}
\and D. Berghmans\inst{\ref{i:rob}}
\and F. Auch\`ere\inst{\ref{i:ias}}
\and R.~Aznar Cuadrado\inst{\ref{i:mps}}
\and L.~Teriaca\inst{\ref{i:mps}}
\and S.K.~Solanki\inst{\ref{i:mps}}
\and C.~Verbeeck\inst{\ref{i:rob}}
\and E.~Kraaikamp\inst{\ref{i:rob}}
\and K.~Stegen\inst{\ref{i:rob}}
\and S.~Gissot\inst{\ref{i:rob}}
}

\institute{Max Planck Institute for Solar System Research, Justus-von-Liebig-Weg 3, 37077 Göttingen, Germany\label{i:mps}
\and
Solar-Terrestrial Centre of Excellence -- SIDC, Royal Observatory of Belgium, Ringlaan -3- Av. Circulaire, 1180 Brussels, Belgium\label{i:rob}
\and
Institut d'Astrophysique Spatiale, CNRS, Univ. Paris-Sud, Universit\'e Paris-Saclay, B\^at.\ 121, 91405 Orsay, France\label{i:ias}
}

\date{Received 11 January 2023 / Accepted 10 July 2023}

 
  \abstract
   {}
   {Within the quiet Sun corona imaged at 1~MK, much of the field of view consists of diffuse emission that appears to lack the spatial structuring that is so evident in coronal loops or bright points. We seek to determine if these diffuse regions are categorically different in terms of their intensity fluctuations and spatial configuration from the more well-studied dynamic coronal features.}
   {We analyze a time series of observations from Solar Orbiter's High Resolution Imager in the Extreme Ultraviolet to quantify the characterization of the diffuse corona at high spatial and temporal resolutions. We then compare this to the dynamic features within the field of view, mainly a coronal bright point.}
   {We find that the diffuse corona lacks visible structuring, such as small embedded loops, and that this is persistent over the 25~min duration of the observation. The intensity fluctuations of the diffuse corona, which are within $\pm 5\%$, are significantly smaller in comparison to the coronal bright point. Yet, the total intensity observed in the diffuse corona is of the same order as the bright point.}
   {It seems inconsistent with our data that the diffuse corona is a composition of small loops or jets or that it is driven by discrete small heating events that follow a power-law-like distribution. We speculate that small-scale processes like MHD turbulence might be energizing the diffuse regions, but at this point we cannot offer a conclusive explanation for the nature of this feature.}

\keywords{Sun: atmosphere --- Sun: corona  --- Sun: UV radiation}

\maketitle
%

\begin{figure*}
 \centering
 \includegraphics[width=165mm]{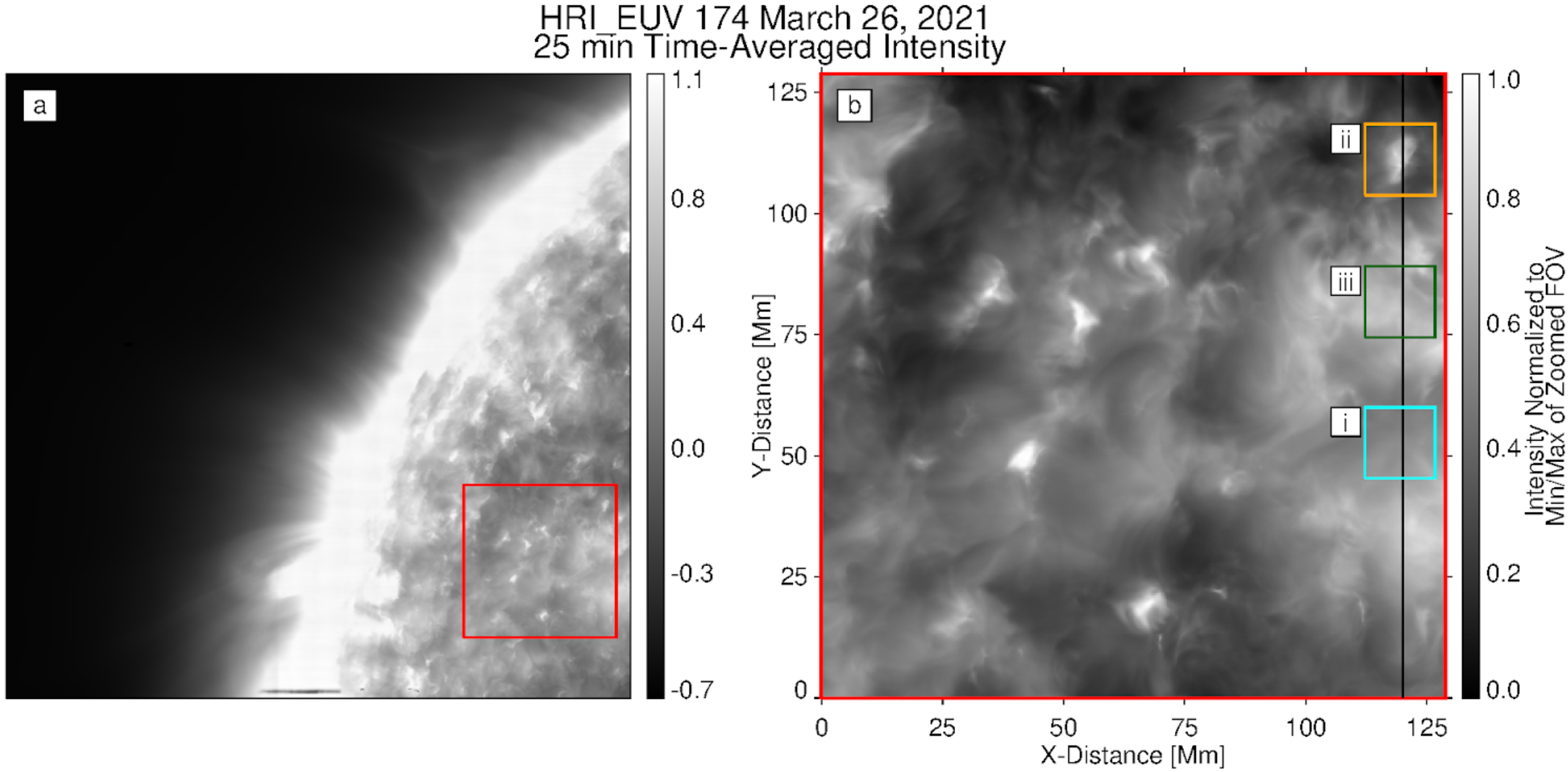}
 \caption{Observation summary. \textit{Panel a}: the full field of view (FOV) covered in the 17.4~nm band of HRI\textsubscript{EUV}. The red box outlines the zoomed-in FOV that is shown in \textit{panel b}. Both \textit{panels a--b} show the time-averaged, normalized intensity  (to the minimum and maximum in the area in panel b). \textit{Panel b}: the vertical black line shows the location of the cut that is used in the time-distance plot displayed in Fig.~\ref{Fig2_tidist} and the intensity-distance plot shown in Fig.~\ref{Fig3_inttime}. The cyan box (i) and the orange box (ii) outline the diffuse region and the bright point used in Fig.~\ref{Fig4_rel_combo}. The green box (iii) outlines the loop-like features zoomed into in Fig.~\ref{Fig.loop.time}. See Sect.~\ref{S.results.imaging}.}
 \label{Fig1_overview}
\end{figure*}

\section{Introduction}
\label{section:Introduction}
In studying the coronal heating problem, three main regions are acknowledged to be observed within the corona. These are active regions (ARs), coronal holes (CHs), and the quiet Sun (QS). Active regions are the brightest and most dynamic portions of the corona, often associated with underlying strong magnetic field patches including sunspots, and seen to consist of coronal loops as observed in the extreme ultraviolet (EUV) and X-rays. In contrast, CHs are the darkest portions in the solar corona in EUV and are attributed to open magnetic fields that connect the solar surface to the heliosphere. Outside of ARs and CHs, there remains the QS. As the QS makes up the largest proportion of the solar surface, understanding the processes at work within this portion of the corona is crucial to understanding coronal heating. 

Based on observations, a variety of small-scale, dynamic features, such as nanoflares, coronal bright points, and jets, have all been considered to be of high importance to balance the overall energy losses from the QS corona \citep[e.g.,][]{aschwanden:2000,hosseini_rad:2021,shen:2021,chitta:2021}. More recently, based on high spatial resolution and high cadence EUV observations from the Extreme Ultraviolet Imager \citep[EUI;][]{rochus:2020} on board the Solar Orbiter \citep[][]{mueller:2020}, \citet{berghmans:2021} observed compact isolated coronal brightenings termed campfires. A main characteristic of all these nanoflare-type heating events, including campfires, is that they are all clearly distinguishable from the local background coronal emission. Whether these observable discrete heating events are sufficient to explain the energy losses from the quiet Sun corona is still an open question \citep[][]{aschwanden:2000,chitta:2021}. While much focus has been placed on these more distinguishable elements or events in the past, there remains much to be learned from the quieter portions of plasma that are devoid of these obvious localized transient brightenings. In particular, we are referring to the areas of seemingly stable and featureless EUV emission in the QS that we label the diffuse corona. 

Diffuse emission associated with ARs has been studied in the past. \citet{viall:2011} looked at the diffuse portions of AR emission (i.e., areas not associated with any distinguishable loops or loop footpoints) and determined that these regions encompass a majority of the emitting portion of ARs, are only marginally less bright (10--35\%) compared to AR loops, and are dynamically heated, as opposed to being energized by some steady process. Whether or not a similar significance and heating mechanism can be attributed to the diffuse corona in the QS remains to be seen.

In most studies, however, this diffuse emission is considered as a background only. Usually, no particular consideration is given to it, apart from the urge to correct for (i.e., subtract) it when looking at features resolved in space and time embedded in this background.

In this work, we analyze the evolution of the diffuse quiescent corona observed with the high spatial and temporal resolution allowed by the EUI instrument. We compare the diffuse emission and its fluctuations to those seen in the more dynamic features within the observational field of view (FOV). We find that not only is the diffuse corona an enduring contributor of seemingly stable and unstructured emission, but that it also is more widespread and is therefore considered to be an important factor in the overall energy balance of the solar corona. 

\section{Observations}
\label{section:Observation}
On March 26, 2021, Solar Orbiter \citep[][]{mueller:2020} was located at a distance of 0.72\,AU from the Sun on the far-side with respect to Earth. The EUV High Resolution Imager (HRI\textsubscript{EUV}) on the Extreme Ultraviolet Imager was pointed towards the limb at a latitude of about 30$^\circ$ and recorded images with very high cadence of 2\,s (1.65\,s exposure) between 23:32:20~UT and 23:57:18~UT (25\,min).\footnote{Data release 4.0 2021-12. DOI: https://doi.org/10.24414/s5da-7e78}

HRI\textsubscript{EUV} has a plate scale of 0.492\arcsec\,pixel$^{-1}$, which amounts to roughly 260\,km\,pixel$^{-1}$ on the Sun (as seen from Solar Orbiter) for this data set. The pass-band of HRI\textsubscript{EUV} is centered at 17.4~nm and its response peaks at temperatures of about 1\,MK due to the presence of spectral lines of Fe\,{\sc ix} (at 17.11\,nm) and Fe\,{\sc x} (at 17.45\,nm\ and 17.72\,nm). Before conducting our analysis, we aligned the level 2 (L2) data to remove the jitter in the image sequence as described in \citet{2022A&A...667A.166C}.

The region observed by EUI is a quiet Sun region outside coronal holes. While the target area was not visible from Earth-based telescopes, the Full Sun Imager (FSI) of EUI acquired images in the 304\,\AA\ and the 174\,\AA\ channels. These show a north polar coronal hole at latitudes well above 60$^\circ$ North, while the field of view of HRI\textsubscript{EUV} is below 55$^\circ$ North, i.e.,\ located far from the coronal hole.

\section{Results}
\label{section:Results}
We focus on the on-disk portion of these observations that also covered the limb and regions off the disk. An overview of the observational FOV, including the primary areas of interest for this study, is shown in Fig.~\ref{Fig1_overview}. Panel a depicts the full-FOV of HRI\textsubscript{EUV}, and panel b shows a zoom into the on-disk portion of the observation that is of interest for this study. For both a and b, the intensity shown has been averaged over the entire 25~min observing time of this EUI sequence, and then it has been normalized by the minimum and maximum intensities within the zoomed FOV shown in b. Thus, the image color scale in the full FOV and the zoom in Fig.~\ref{Fig1_overview} is different. With the above normalization for panel b it is limited between 0 and 1 and  for panel a, which contains pixels with both greater and weaker intensities, it ranges from $-$0.7 to 1.1. 

\subsection{Diffuse region, loop-like features, and coronal bright points\label{S.results.imaging}}

We find several distinct regions in the corona at $\sim$1\,MK. There are (1) dark patches (with normalized intensities of 0.25 or less), (2) bright patches associated with loop-like structures and coronal bright points containing intensities at or near the saturation limit of the detector, and, finally, (3) there appear to be what we refer to as ``diffuse'' regions.

These diffuse regions are areas that seem to lack structuring in both time and space. They are hazy portions seen throughout the FOV with time-averaged, normalized intensities mostly between 0.25--0.6. An example patch of diffuse corona is outlined by the cyan box (panel i) in Fig.~\ref{Fig1_overview}b. 
The diffuse corona appears to cover a significant area when compared to the brighter regions in the FOV. However, the diffuse corona lacks any of the obvious structuring that is associated with the well-studied structures of the quiet solar corona, such as coronal bright points that are composed of $\sim$10\,Mm long loops \citep[e.g., see the review article on coronal bright points by][]{madjarska:2019}.

\begin{figure}
 \centering
 \includegraphics[width=88mm]{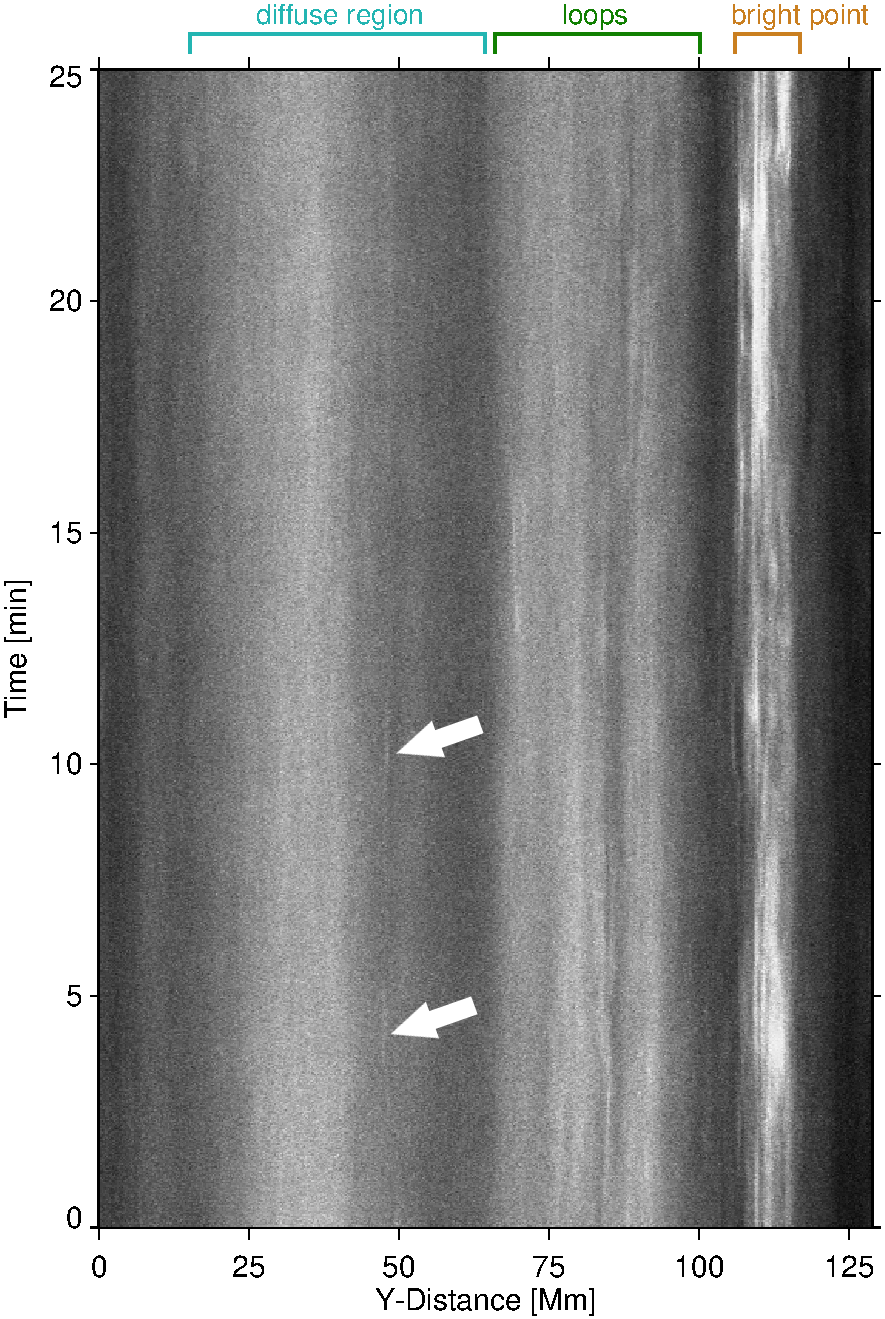}
 \caption{Temporal evolution of cut through diffuse region and coronal bright point. This time-distance plot shows the intensity in HRI\textsubscript{EUV} from along the cut (1 pixel in $x$) outlined by the black vertical line in Fig.~\ref{Fig1_overview}b as it evolves over the entire 25~min observing time. The intensity is scaled linearly from 915~DN\,s$^{-1}$ (black) to 2500~DN\,s$^{-1}$ (white). Above the panel the location in the $y$ direction of three types of regions are marked. The two arrows mark transient brightenings in the diffuse region. See Sects.~\ref{S.results.imaging}, and \ref{S:loops}.
 \label{Fig2_tidist}}
\end{figure}

Our analysis aims to compare these differing regions in both a qualitative and quantitative manner to determine the extent of this apparent discrepancy in terms of spatial structuring. For this we investigate a cut through the FOV that runs through these different features (black line in Fig.~\ref{Fig1_overview}b): diffuse regions, loop-like features, and a coronal bright point. (While many bright features are saturated in the exposure, in particular at the limb, this bright point is not). The intensity along this cut as a function of time is presented in a time-distance plot (see Fig.~\ref{Fig2_tidist}). It is evident from this plot that the bright point stretching from $y{=}$106~Mm to 117~Mm shows intensity structuring in both space and time. The same also applies to the clear loop-like features situated around $y{=}$65~Mm to 100~Mm. The bright point and the loop-like features appear to be continually evolving on spatial scales of roughly a few Mm and timescales on the order of a few minutes, although the loop-like features are comparatively less variable than the bright point.

\begin{figure}
 \centering
 \includegraphics[width=88mm]{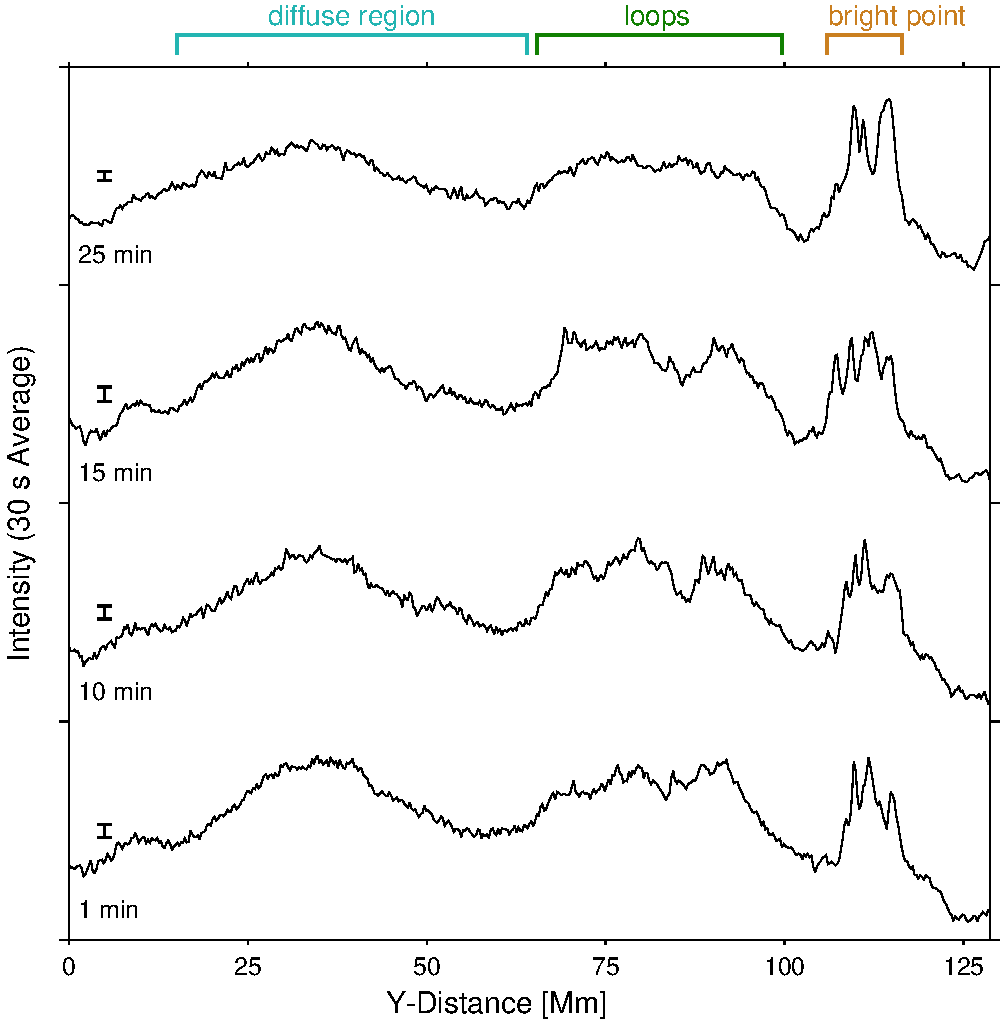}
 \caption{Spatial variability along a cut through a diffuse region, loop-like features and a coronal bright point at four different times. This intensity-distance plot shows the intensity along the cut outlined by the black vertical line in Fig.~\ref{Fig1_overview}b for several intervals of time during the observation. For each 30~s interval (15 time-consecutive images), the intensity at each pixel along the cut is averaged over that time period and plotted as a function of distance. Four of these light curves are shown as a stack, with the bottom curve being the intensity-distance 30~s average ending at 1~min (and, therefore, starting at 30~s), the curve directly above it is the 30~s average ending at 10~min, and so on. The black vertical bars with each curve show the maximum error for that curve (see Appendix~\ref{S:errors}). Above the panel the location in the $y$ direction of three types of regions are marked. See Sects.~\ref{S.results.variability} and \ref{subsection:Discussion_EmissionContr}.
 \label{Fig3_inttime}}
\end{figure}

\begin{figure*}
 \centering
 \includegraphics[width=140mm]{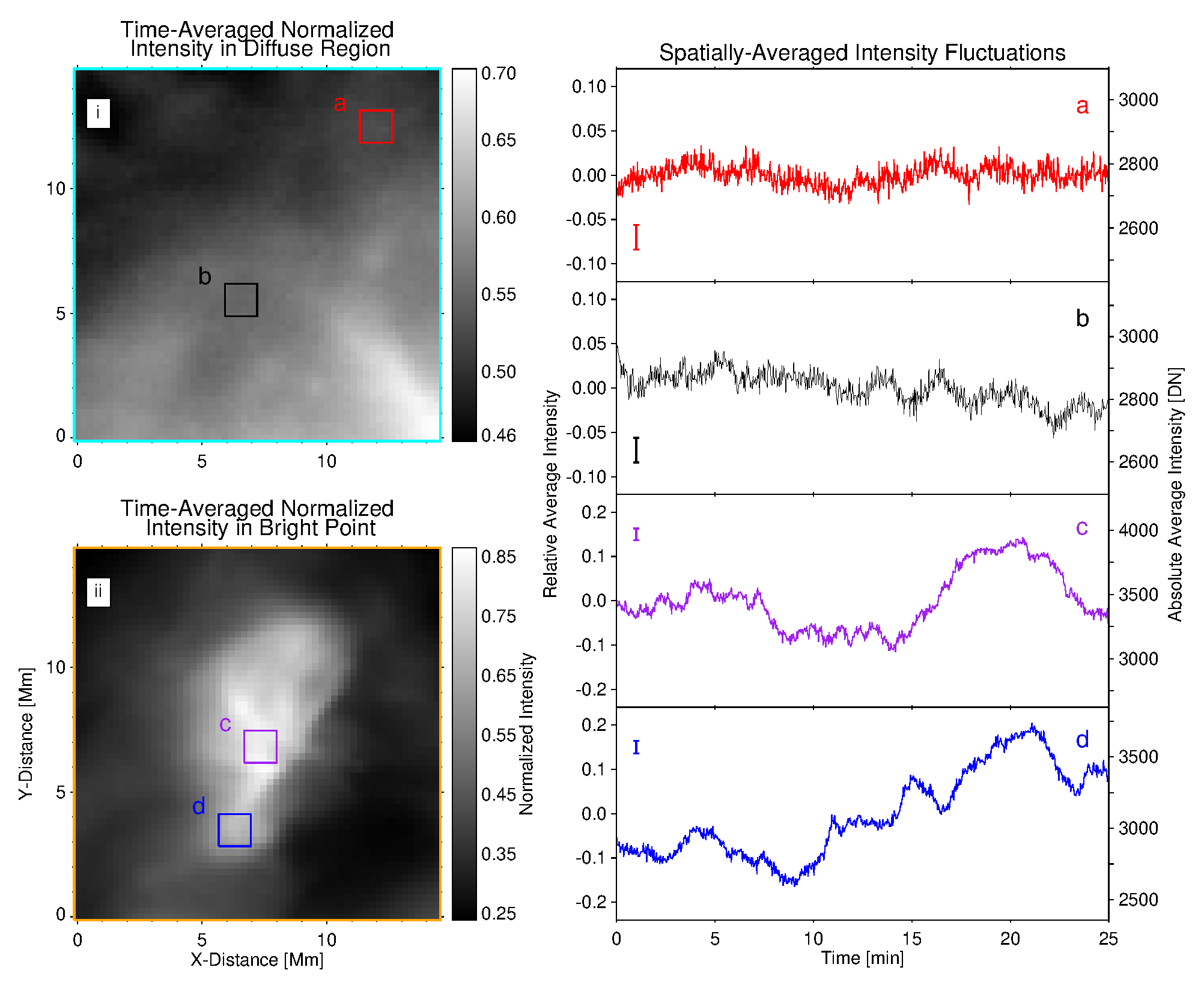} 
 \caption{Intensity fluctuations in diffuse and bright point regions. The two left panels show the zoomed-in FOV covering the diffuse region (panel i) and bright point (panel ii) outlined by the cyan and orange boxes in Fig.~\ref{Fig1_overview}b, respectively. Each image is the same time-averaged and normalized figure as shown in Fig.~\ref{Fig1_overview}b, except the intensity range shown is further limited to the minimum and maximum for each zoomed FOV. Within the regions of interest, four smaller sub-fields a to d are outlined. Each of these boxes has the same area. The spatially-averaged relative (left y-axis) and absolute (right y-axis) intensities for the respective sub-fields are plotted as a function of time on the right in panels a--d. The vertical bars shown in each of these panels represent the maximum error for each time-series (see  Appendix~\ref{S:errors}).} 
 \label{Fig4_rel_combo}
\end{figure*}

The bright point is a typical small coronal bright point in appearance and size. It consists of a number of short coronal loops with the core region (Fig.~\ref{Fig4_rel_combo}\,ii) having a size of just over 5~Mm \cite[cf.][]{madjarska:2019}. Such compact small-loop-type bright points are found in abundance in the quiet Sun and have been used, e.g., to determine the coronal rotation \cite[][]{2001A&A...374..309B,2002A&A...392..329B}, even though the spatial resolution of the older data was inadequate to sufficiently resolve the internal structure of the bright points.
The region with loop-like features does not show clearly distinguishable loops, but (in the time-averaged image) elongated features reminiscent of loops (Fig.~\ref{Fig.loop.time}\,iii). Because we concentrate in this study on the diffuse regions, we do not follow this up further (see Appendix \ref{S:loop-like} for a further discussion of the loop-like features).

In contrast to these more dynamic features, diffuse regions seen in Fig.~\ref{Fig2_tidist} are rather invariable. The time-distance plot reveals the overall temporal stability of the diffuse corona. The emission is diffuse in the $y$-direction in a range of 15--65~Mm, with two seemingly distinct regions of differing intensity levels: a brighter portion between 15--42~Mm and a darker portion from 42--65~Mm. For both portions, there is a hazy quality that remains stable throughout the 25~min observing time on spatial scales comparable to that of a supergranule. Both the brighter and darker sections show no obvious jumps in intensity across their width. There is a small brightening at about $y{=}$48~Mm occurring periodically over intervals of time limited to a few minutes (highlighted by arrows in Fig.~\ref{Fig2_tidist}). This could be categorized as an HRI\textsubscript{EUV} campfire \citep[see][]{berghmans:2021}. However, the brightening is only a few pixels in width and makes up only a small fraction of the overall emission in that region.

We emphasize that the data shown in the space-time plot in Fig.~\ref{Fig2_tidist} have a time cadence of only 2~s and a spatial sampling of 260~km on the Sun. This implies that transient small-scale features should be visible if they were present (down to those temporal and spatial scales).

\subsection{Spatial and temporal variability\label{S.results.variability}}

We now dissect the time-distance plot for a closer look at the difference between the coronal bright points and the diffuse regions. To this end, the intensity along the cut from Fig.~\ref{Fig1_overview}b was averaged over each subsequent 30~s interval and displayed as line plots of time-averaged intensity versus distance, several of which are shown stacked in Fig.~\ref{Fig3_inttime}. Essentially these stacked plots are horizontal cuts through Fig.~\ref{Fig2_tidist} for a given time span.

Across the aforementioned diffuse regions (i.e., $y{=}$15--65~Mm), spatial changes in the intensity are gradual. As will be discussed below, the small fluctuations seen on top of this gradual variation are at the level of the calculated maximum error. The shape of the curve also remains similar for each plot, despite there being 5--10~min of separation between them. This implies that the diffuse corona behaves in a spatially coherent fashion over its entire extent corresponding to the scale of a supergranule of about 20~Mm.

The peak intensity in the brighter diffuse area (at $y{\approx}35$~Mm) is only slightly less than, or even equal to the peak intensity seen in the bright point and loop-like features (in particular before $t{=}15$~min). Conversely, the morphology of the loop-like features and coronal bright point changes markedly over time, and the intensity fluctuations within these features at each point in time varies significantly, i.e., well-above the maximum error. This further differentiates the characteristics of the diffuse regions from those of the bright points and loop-like features.

To quantitatively characterize the temporal stability of the diffuse corona, its relative intensity fluctuations provide clearer insight (Fig.~\ref{Fig4_rel_combo}). Here, we compared the absolute and relative intensity variations of the diffuse corona (box i in Fig.~\ref{Fig1_overview}b) to that within the coronal bright point (box ii). The absolute fluctuations are derived by simply spatially averaging the intensity within each of the sub-fields a to d for each snapshot. The relative fluctuations were then calculated by subtracting the overall time-averaged intensity within each sub-field from its spatially-averaged intensity and then dividing this difference by the time-averaged intensity. Figure~\ref{Fig4_rel_combo}i and ii show which sub-fields of the observation are sampled, either from a diffuse region (a, b) or from a bright point (c, d). 

The relative fluctuations in the sub-fields of the diffuse region (a, b) remain within a few percentage points (less than $\pm$5\%) over the entire 25~min (Fig.~\ref{Fig4_rel_combo}a, b). In contrast, in the sub-fields of the coronal bright point (c, d), the intensity changes by over 10\% on shorter timescales of about 5~min and can sway by up to almost 40\% over the entire 25~min (Fig.~\ref{Fig4_rel_combo}c, d). 

To judge the significance of these fluctuations, one has to compare the observed variability to the measurement errors. In Appendix \ref{S:errors} we discuss how we calculate the maximum estimated error. As a very rough estimate of this, for our current data set in regions that are moderately bright, the typical error of the intensity is of the order of 3\%.

In general, the sub-fields of the bright point show trends that rise well above the level of the maximum estimated error, both on shorter and longer timescales. In contrast, in the diffuse regions the variability we see would essentially be consistent with measurement errors. There might be some dynamics within the diffuse regions with amplitudes just above the noise that persist for more than a few time steps, but these fluctuations are still smaller than the overall level of fluctuations over the observing period (of 25 min). While the fluctuations shown here are only for two sub-sections of each feature, the behavior is similar when more sub-sections are analyzed.

\begin{figure*}
 \centering
 \includegraphics[width=140mm]{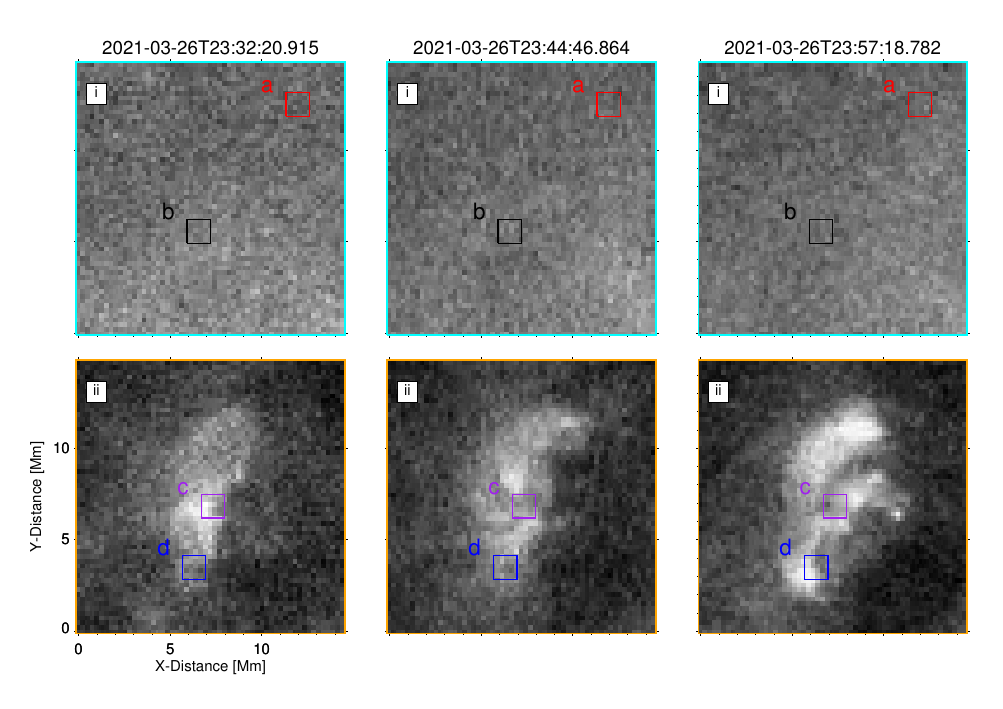} 
 \caption{Snapshots of diffuse region and bright point without time-averaging. The snapshots at three times as indicated by the time stamps are shown in the three columns. The top row shows the the diffuse region (marked i) and the bottom row the bright point (marked ii). Images are commonly scaled to the same minimum and maximum values of intensity. Boxes a--d have the same meaning as in Fig.~\ref{Fig4_rel_combo}. See Sects.~\ref{S.results.variability} and \ref{subsection:Discussion_EmissionContr}.} 
 \label{FigA2}
\end{figure*}

This relative temporal stability of the diffuse corona and the larger variability seen in the coronal bright point are further demonstrated using the respective snapshots, at three instances, without time-averaging (see Fig.~\ref{FigA2}). It is clear from these images that the diffuse region is not only temporally stable compared to the bright point, but it also lacks any distinguishable spatial structuring.

This time series analysis underlines that the diffuse corona, often considered as a mere background, contributes a significant amount of the coronal emission in the quiet Sun. For a quantitative estimate of this contribution, see Sect.~\ref{subsection:Discussion_EmissionContr}.

\section{Discussion}
\label{section:Discussion}
We note the general pervasiveness of the diffuse corona seen at temperatures of 1~MK with HRI\textsubscript{EUV}. These regions remain featureless in both time and space, yet are relatively bright compared to the more dynamic features within the FOV. This naturally leads to questions about the significance of these regions in terms of their contribution to the emission of the 1~MK corona and on the possible heating mechanisms operating in such regions. Similarly, questions arise regarding the reasoning behind the lack of noticeable structuring and how this is related to the overall energy balance within the solar atmosphere.

\subsection{Contribution of diffuse regions to radiative losses}
\label{subsection:Discussion_EmissionContr}
The intensity observed in the diffuse regions is relatively strong, even when compared with bright points. This is illustrated in Fig.~\ref{Fig4_rel_combo}a-d where we show the relative and absolute average intensity values [DN] within sub-fields of the diffuse and bright point regions. We see that, over the entire observing period, the time-averaged diffuse intensity (i.e., radiative flux per area) is about 80\% that of the bright point, i.e., of the same order of magnitude. Also, from Fig.~\ref{Fig3_inttime}, the peak intensities along the cut from the brighter diffuse portion (around $y=\,$35~Mm) are very similar to those from the bright point (around $y=\,$110~Mm) and also about equal to those from the loop-like features sampled between $y=\,$65~Mm and 100~Mm. All of this points to the conclusion that the emission coming from the diffuse regions is non-negligible and should be an important consideration in any coronal heating model, in particular when considering that the diffuse corona can cover a large fraction of the quiet Sun.

\begin{figure}
 \centering
 \includegraphics[width=80mm]{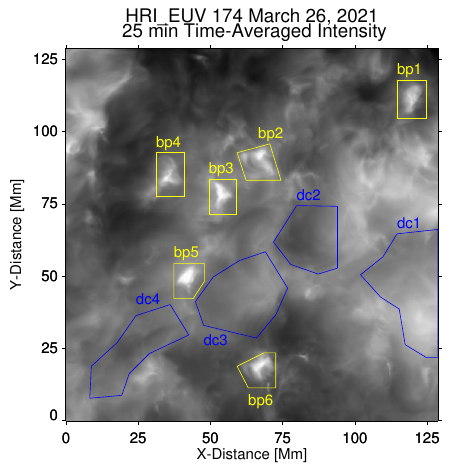}
 \caption{Grouping of diffuse coronal regions and bright points in quiet Sun. The intensity image is the same as Fig.~\ref{Fig1_overview}b. The diffuse coronal regions (dc1 to dc4) are outlined in blue, the bright points (bp1 to bp6) are highlighted in yellow. See Sect.~\ref{subsection:Discussion_EmissionContr}.}
 \label{FigA1_diff_vs_bps}
\end{figure}

We further investigate the significance of diffuse emission by conducting a rough estimate of the total emission contribution from the diffuse areas compared to the bright points within the FOV. 
For this we classify the regions in the FOV in Fig.~\ref{Fig1_overview}b into areas covered by coronal bright points and diffuse regions by a simple by-eye estimate. These regions are marked and labeled in Fig.~\ref{FigA1_diff_vs_bps}. Some of the bright points host a few pixels that are saturated on the detector. Hence our estimate for the bright point emission will be a lower limit only. However, this effect should be well below a  factor of two and thus our calculation should be just fine for the order-of-magnitude estimate we aim for here.

Based on the integration of the emission from the respective diffuse and bright point areas, we find that the seemingly quiet diffuse areas provide almost 2.7 times the emission as the more dynamic features. Thus overall, the diffuse emission would dominate the quiet Sun radiative losses at around 1\,MK, while discernible bright features would contribute a minor fraction, maybe half at best. This is not surprising since these by-eye defined diffuse regions also have almost three times the area as the bright points (see Fig.~\ref{FigA1_diff_vs_bps}).

\subsection{Diffuse quiet-Sun corona and small loops\label{S:loops}}

The magnetic field is space-filling and is the driver behind the energetics of the corona. The work of \citet{dowdy:1986} described the solar magnetic scene to be comprised of both (locally) open field lines expanding into funnels with height and small-scale, closed loops that connect back to the surface. Based on studies of magnetic field extrapolations, such loops are expected to be rooted not only in the network regions at the edges of supergranules, but also in the internetwork within a few Mm of the network boundaries \citep{schrijver:2003,2010ApJ...723L.185W}.

Quiet Sun observations reveal the presence of very short loops clearly distinguishable in EUV observations, only a few Mm long. They come at (probably) high coronal temperatures \cite[][]{2013A&A...556A.104P,2017A&A...599A.137B} as well as at lower transition region temperatures \cite[][]{2014Sci...346E.315H}. These short loops have lifetimes of only a (few) minute(s). More recent observations also reveal the dynamic nature and substructure of such small EUV loops including propagating features \cite[][]{2021A&A...656L..16M} and small jets associated with them \cite[][]{2021A&A...656L..13C}. 
Also, transition region loops crossing a super-granular cell in the quiet Sun (i.e., with a length of ca. 20~Mm), have been reported \cite[e.g., Fig.~1 of][]{2004A&A...427.1065T}. %
However, all these types of loops are discrete units and far from space-filling.

Still, the presence of loops in the quiet Sun would mean that we should expect the presence of a few Mm long, small-scale coronal loops in our observation even within the so-called diffuse areas. As discussed above, these smaller loops are expected to be dynamic, showing brightness fluctuations on timescales on the order of minutes \citep[see also][]{reale:2010}.
Similar lifetimes of (magnetic) loops of a few minutes are also found based on magnetic field extrapolations from time series of high resolution quiet Sun magnetograms \citep[][]{2010ApJ...723L.185W}.
Case in point, our observations do show such brightness fluctuations in the coronal bright points, which are indeed on shorter timescales compared to the 25~min of observed stability for our diffuse areas.

If the diffuse areas are composed of such smaller loops, then we would expect loop-like brightenings at some point during our analysis unless a different loop emission behavior is at play. We do see that there are a limited number of small-scale brightenings occurring within some diffuse portions of Fig.~\ref{Fig2_tidist} (see white arrows) that last for a few minutes at a time and are only a few pixels in length (i.e., some 500~km in $y$-distance). These brightenings, however, are not ubiquitous as would be expected for the typical picture of low-lying loops crisscrossing everywhere in the FOV. Also, these small intensity enhancements barely rise above the level of the local diffuse region (or background). When compared to the intensity level local to the area of the brightening in the minutes before and after its enhancement, the brightening only rises to a level of 5\% above this background.

The coherent spatial intensity variations on supergranular scales in these diffuse regions indicate that all the constituting smaller loops, if present, must evolve in unison, which is unlikely.
Based on this, we think that the small (coronal) loops reported before may not be the source of emission from the diffuse regions. In that case we should (occasionally) see a transient brightening caused by one of the transient Mm-scale loops that have been reported before, but we do not see these.

\subsection{Diffuse quiet-Sun corona and jets\label{S:spicules}}

Chromospheric spicules \citep{depontieu:2007} and transition region network jets \citep[][]{2014Sci...346A.315T} are other common, small-scale jet features whose imprint should be seen everywhere on the quiet Sun. Spicules are chromospheric jets and are categorized into two types \citep{depontieu:2007}. Type-I spicules are longer-lived and seen to both rise and fall at the limb, remaining at chromospheric temperatures throughout their lifetime. Type-II spicules are more impetuous and often appear to shoot up before disappearing from chromospheric imaging channels, sometimes then appearing in the hotter channels imaging transition region (TR) plasma \citep{pereira:2014}, which can also include the return flows when cooling back down from hotter temperatures \cite[][]{2021A&A...654A..51B}. Spicules have typical speeds ranging from tens of ~km\,s$^{-1}$ (type-I) up to 100~km\,s$^{-1}$ (type-II), lifetimes of several minutes, and characteristic lengths ranging from a few hundred to several thousand km \citep[e.g.,][and included references]{tsiropoula:2012}.

While spicular material is easiest to detect at the cooler temperatures found in the chromosphere and TR (T~$\le 10^5$~K), recent investigations show that there are exceptions to this where spicule signatures can be observed as corrugations in the EUV emission. For example, signatures have been found that the on-disk counterparts of spicules, namely dynamic fibrils, show EUV emission \cite[][]{2023A&A...670L...3M}. This might indicate higher temperatures of more than $10^5$\,K, although this is not conclusive, yet \citep{henriques:2016,martinez-sykora:2018,samanta:2019}. Similarly, coronal counterparts of network jets remain elusive \citep[][]{2018A&A...616A..99K,2022A&A...660A.116G}.

Small coronal jets are, in general, a common phenomenon in quiet Sun regions. In particular, recent EUI observations have shown an abundance of small-scale jets \cite[e.g., ][]{2021A&A...656L..13C,2022A&A...664A..28M}. Small jets and their substructure, e.g., plasmoids, have recently also been observed in radio emission \cite[e.g.][]{2017ApJ...841L...5S,2019ApJ...875..163R}.
Still, the question remains if highly dynamic small-scale features such as these jets or spicules can provide an explanation of the diffuse corona.

At 2~s cadence and about 500~km resolution, the observation from HRI\textsubscript{EUV} that we analyze in this study certainly could pick up any spicules or network jets that would show counterparts in EUV. Yet, we do not detect any such signatures in the far-reaching diffuse areas. This either means that these jets only very rarely reach coronal temperatures, or at least do not get heated to 1~MK in the magnetic regime that is responsible for the diffuse emission. Another possibility could be that those small-scale jets that do reach coronal temperatures might lose their identity and structuring as their energy gets dissipated.

\subsection{Diffuse quiet-Sun corona and small-scale heating events}
\label{subsection:Discussion_HeatMechs}

The simple presence of a diffuse, seemingly featureless structure has implications for the heating mechanism that has to sustain the hot corona. The energization has to be either continuous or be concentrated in a very large number of small-scale heating events so that this is not leaving an imprint at the resolution of our observations. Hence, we explore the applicability of several proposed heating mechanisms for the diffuse corona.
Clearly, we cannot relate our findings to all possible heating scenarios, instead we picked those we considered most relevant for our study. The reader is referred to recent reviews that discuss the heating problem in general \cite[e.g.,][]{2006SoPh..234...41K}, with respect to 3D models
\cite[e.g.,][]{2015RSPTA.37350055P}, in terms of waves \cite[e.g.,][]{2020SSRv..216..140V}, or in the light of field-line braiding \cite[e.g.,][]{2020LRSP...17....5P}.

One heating scenario involves a power-law distribution of the number of brightening events with energy characterized by a slope of $-$2, at least for the smaller flare-like events with less than $10^{26}$~ergs \citep[][]{hudson:1991}. This requires there to be an increasingly higher number of distinct events at ever smaller spatial and energy scales. In fact, these events would be so small that they have not yet been resolvable by Sun-observing instruments \citep{parnell:2000,chitta:2021}. Up until now, however, results are mixed regarding whether or not the energy distribution for impulsive heating events has the necessary slope to account for coronal heating \citep[e.g.,][]{berghmans:1998,krucker:1998,aschwanden:2000,parnell:2000,pauluhn:2007,aschwanden:2013}. It has also been pointed out that even if the power-index is greater than 2, this is not necessarily sufficient to heat the quiet Sun \citep{berghmans:2002}. \citet{joulin:2016} argue that, regardless of the actual energy distribution slope, it is inherently unlikely to be able to detect high-frequency, small-scale brightenings against the background coronal emission. The authors state that their study combined with the works of others before them shows that heating cannot be guaranteed to be observed as broken up into discrete events.

Perhaps such unresolved, impulsive brightenings are what is the cause behind the diffuse corona. If there were a sufficiently high number of small-scale events resulting in the diffuse emission that we see, they would still have to be driven in a spatially coherent fashion to create a diffuse region about 10 to 20~Mm large in size, i.e., a region the size of a supergranule. Considering the significant structuring found at the base below the corona, in the chromosphere, it is not very plausible that the diffuse corona is driven by small-scale events.

Should small-scale events drive and heat the diffuse coronal patches, one would also expect larger events in these regions. This is based on the finding that even on the smallest scales resolved so far power-law-like distributions prevail \cite[][]{berghmans:2021}. However, with two exceptions of events barely resolved by HRI\textsubscript{EUV} (arrows in Fig.~\ref{Fig2_tidist}), we do not see any distinguishable transient brightenings in the diffuse regions (above the noise level). A more rigorous statistical analysis is required to draw a final conclusion, but at this point we consider it unlikely that individual events distributed through a power law could create a diffuse corona as we observe here.

One might speculate that some form of MHD turbulence might lead to energy dissipation far below the scales resolvable by current observations and by this create a quasi-continuous heating in these regions. Of course, this would raise the question why this should be operating in the diffuse regions while other areas show much higher contrast in the coronal quiet Sun features. As such a discussion would go far beyond this observational study, we refrain from any further speculation on this.

\subsection{Diffuse quiet-Sun corona and wave heating}
\label{subsection:Discussion_Wave}

If the energization of the hot, diffuse quiet-Sun corona has to be quasi-continuous, then wave heating could also be an option. In the first theoretical attempts to explain the hot outer atmosphere, heating through waves was already suggested, at that time through the dissipation of sound waves \cite[][]{Biermann1946,1948ApJ...107....1S}. On average, an energy flux of about 100~W\,m$^{-2}$ is required to balance the energy losses of the corona in the quiet Sun \cite[e.g.,][]{1977ARA&A..15..363W}, and in general upward propagating magneto-acoustic waves have the potential to heat the plasma in magnetically closed structures \cite[e.g., review by][]{2015RSPTA.37340261A}. Such waves have been observed \cite[e.g.,][]{2007Sci...317.1192T} and they carry an energy flux sufficient to heat the quiet Sun corona \cite[e.g.,][]{2011Natur.475..477M}.

Assuming that the upward-propagating waves are generated by p-mode leakage \cite[e.g.,][]{2009SSRv..149...65D}, one might expect quite a homogeneous distribution of the energy flux into the upper atmosphere. While in the photosphere we can expect a spatial structure on the scale of granulation, the rapid expansion of the magnetic field with height guiding the wave flux might quickly even out spatial inhomogeneities. 
However, a detailed investigation of the expansion of the underlying magnetic field in the diffuse coronal regions would be required before drawing any final conclusions on this.

In the time series of small sub-fields, some indications for 3-minute oscillations might be found. A by-eye inspection of Fig.~\ref{Fig4_rel_combo}a-d suggests the presence of fluctuations on a time scale of a few minutes in the diffuse corona (panels a,b) and the bright point (c,d). If such fluctuations would be present, that would indeed be supportive of the leakage of wave power from the photosphere into the higher atmospheric regions seen in the diffuse corona. Just as with the expansion of the magnetic field, a detailed analysis of a possible presence of 3-minute oscillations in the diffuse regions will have to be conducted in the future.

\section{Conclusions}
We analyze the diffuse quiescent corona observed by HRI\textsubscript{EUV} on board Solar Orbiter. Here we find large patches of diffuse corona lacking resolvable physical structuring on  scales below those of a supergranule, i.e., about 20~Mm. Still, the coronal emission from these diffuse regions is of comparable brightness to the more dynamic features like loop-like features or coronal bright points. The spatial variability in the diffuse corona is below about 5\%. The diffuse corona remains temporally stable throughout the observing period, i.e., for at least 25 minutes.

This diffuse regime is rather commonplace within the coronal makeup, contributing a large proportion of the emission seen at 1~MK, yet its underlying nature is still unclear. We consider it unlikely to be connected to features such as spicules or small loops. A power-law-like distribution of discrete heating events seems inconsistent with our observations. We speculate that small-scale processes like MHD turbulence or upward-propagating waves might be energizing the diffuse regions, but at this point we cannot offer a conclusive explanation for the nature of the diffuse regions.

The lower atmosphere in the quiet Sun shows a high degree of temporal and spatial complexity. This is illustrated, e.g., by the cartoon picture suggested by \cite{2009SSRv..144..317W} in their Fig.~16. The diffuse region we investigated in this study might well be related to the mixture of structures in the inter-network, above and below the canopy domain. This would also fit into the interpretation of \cite{2023A&A...673A..81M} of a diffuse region between network patches of the same polarity on opposite sides of a super-granular cell.

A more extensive investigation of this diffuse component of the quiet Sun's corona, using more and more diverse datasets, would help to better estimate how common this component is, including observations of plasma at other temperatures. Also, the brightness distribution in the diffuse patches, the lifetime and size distributions of such diffuse patches, and if wave patterns (of whatever nature) are commonly seen in them needs further consideration.
Just as important will be studies combining EUV data with magnetic field measurements, e.g., provided by the SO/PHI instrument \cite{2020A&A...642A..11S}.
Finally, there is also a clear need for studies of possible heating mechanisms leading to such diffuse parts of the corona.

\begin{acknowledgments}
This work was supported by the International Max-Planck Research School (IMPRS) for Solar System Science at the University of G\"{o}ttingen. L.P.C. gratefully acknowledges funding by the European Union (ERC, ORIGIN, 101039844). Views and opinions expressed are however those of the author(s) only and do not necessarily reflect those of the European Union or the European Research Council. Neither the European Union nor the granting authority can be held responsible for them. Solar Orbiter is a mission of international cooperation between ESA and NASA, operated by ESA. The EUI instrument was built by CSL, IAS, MPS, MSSL/UCL, PMOD/WRC, ROB, LCF/IO with funding from the Belgian Federal Science Policy Office (BELSPO/PRODEX PEA 4000134088, 4000112292, 4000117262, and 4000134474); the Centre National d’Etudes Spatiales (CNES); the UK Space Agency (UKSA); the Bundesministerium für Wirtschaft und Energie (BMWi) through the Deutsches Zentrum für Luft- und Raumfahrt (DLR); and the Swiss Space Office (SSO).
\end{acknowledgments}


\begin{appendix}

\section{Preliminary error estimates\label{S:errors}}

To better quantify and compare intensity fluctuations, we calculated the total error, $\sigma_{\rm{total}}$, associated with the observed emission in the following manner. First, we converted the observed, spatially averaged intensities, $I$, from DN\,s$^{-1}$ to DN by multiplying them by exposure time, $t_{\rm{exp}}$, and then converted this to photon count by considering the current high gain estimate, $a$, for the HRI\textsubscript{EUV} channel. Next, we used this to calculate the photon noise, $\sigma_{\rm{ph}}$, by taking the square root of the total number of photons. This photon noise was then divided by the square root of the number of pixel values, $N$, averaged over to obtain the initial average intensity. After converting the photon noise back into DN with the conversion factor, $a$, the square of this amount was summed with the square of the read-out noise, $R$, and then the square root of this sum gave the final $\sigma_{\rm{total}}$. A final, reduced equation for this is:

    \begin{equation}
        \sigma_{\rm{total}} = \sqrt{R^2+\frac{I\,t_{\rm{exp}}\,a}{N}} \,,
    \end{equation}

Here, $\sigma_{\rm{total}}$ is in DN, $R=\,$2~DN, $I$ is in DN\,s$^{-1}$, $t_{\rm{exp}}$ is in s, and $a=\,$6.85~DN\,photon$^{-1}$. For a typical bright point with an average intensity of 3450 DN, we find a maximum error of 27 DN. For the diffuse regions, a typical average intensity of 2765 DN has a maximum error of 23 DN. Due to the current early stage of the Solar Orbiter mission, the error estimation calculation is not yet fully clarified. So, we would expect that any reasonable calculated errors should be comparable to the root-mean-square (RMS) fluctuations, as these are expected as an upper limit. For this observation, the maximum errors estimated in the manner detailed above are comparable to the high-frequency (1~min) RMS fluctuations which are of a typical order of a few percent. Thus, we consider our error estimation to be reasonable. 
%

\section{Region of loop-like features\label{S:loop-like}}

In the main text we concentrated on the properties of the diffuse region and compared these to the coronal bright point.
Here we concentrate on an area with loop-like features (green box iii in Fig.~\ref{Fig1_overview}). This region shows a spatio-temporal structuring of a higher degree than the diffuse region but much less variability than the bright point (see Fig.~\ref{Fig2_tidist}). In this appendix we briefly compare the properties of this region with loop-like features to the diffuse region and the bright point.

\begin{figure*}
 \centering
 \includegraphics[width=140mm]{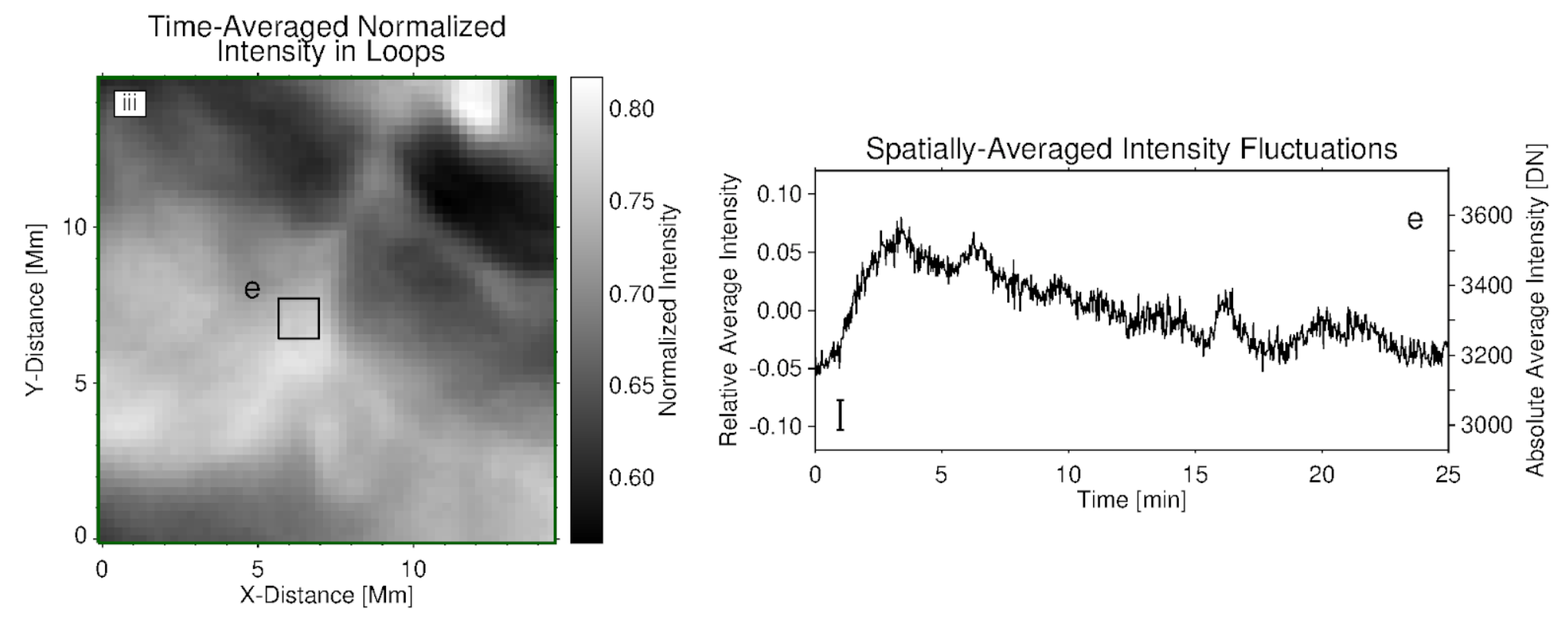} 
 \caption{Intensity fluctuations in the region with loop-like features. The layout of this figure corresponds to Fig.~\ref{Fig4_rel_combo}.
 The left panel shows the zoom into the green box in Fig.~\ref{Fig1_overview}b (region iii). Within this region a smaller sub-field e is outlined. This has the same area as the sub-fields in Fig.~\ref{Fig4_rel_combo} i and ii. The spatially-averaged relative (left y-axis) and absolute (right y-axis) intensity for sub-field e is plotted as a function of time in the right panel. The vertical bar represents the maximum error for the time-series. See Appendix~\ref{S:loop-like}.}
 \label{Fig.loop.time}
\end{figure*}

The time averaged intensity map of this region shows distinct structures, even though it is not simple to attribute a particular nature to these features. The left panel of Fig.~\ref{Fig.loop.time} shows a zoomed view of this area. In contrast to the diffuse region (Fig.~\ref{Fig4_rel_combo}i) it shows clear elongated structures. For simplicity we call these loop-like features. 

The fluctuations seen in this region are comparable or larger than in the diffuse region but much less pronounced than in the bright point. This is evident when comparing the time variability near the center of the loop-like features (Fig.~\ref{Fig.loop.time}e) to the corresponding variability seen in the other regions (Fig.~\ref{Fig4_rel_combo}a-d). 

\begin{figure*}
 \centering
 \includegraphics[width=140mm]{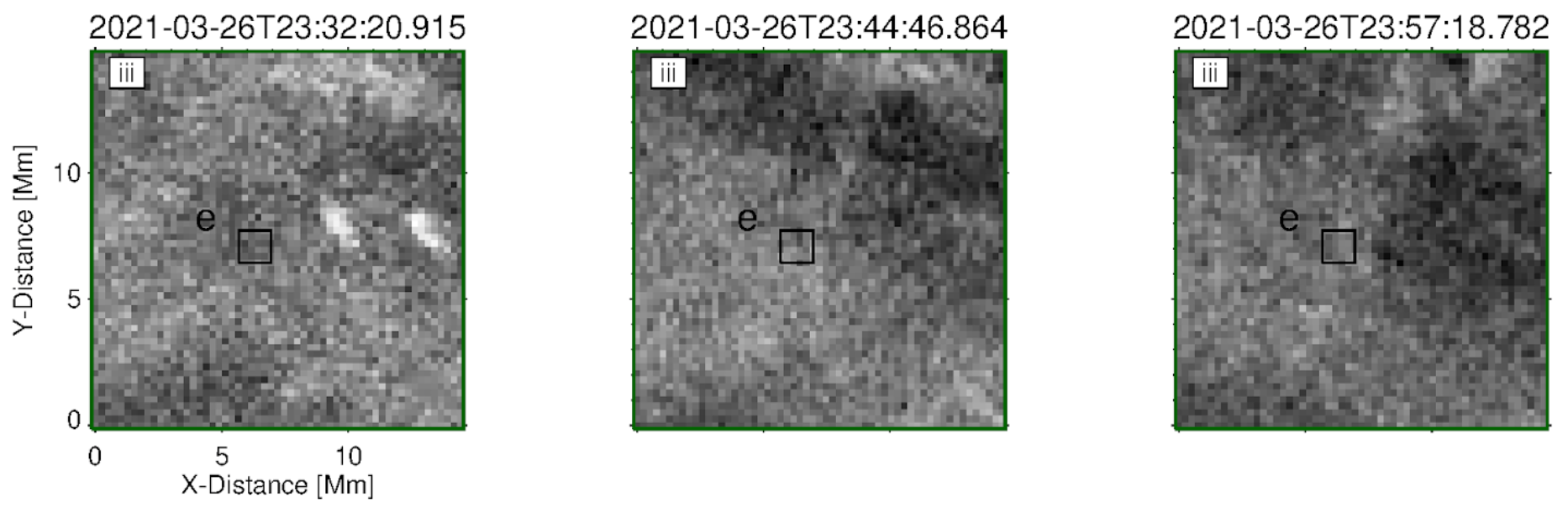} 
 \caption{Snapshots of region with loop-like features without time-averaging. 
 This figure corresponds to Fig.~\ref{FigA2}.
 The snapshots are shown at three times as indicated by the time stamps. Images are commonly scaled to the same minimum and maximum values of intensity. Box e has the same meaning as in Fig.~\ref{Fig.loop.time}. 
 See Appendix~\ref{S:loop-like}.}
 \label{Fig.loop.snap}
\end{figure*}

The individual snapshots in this region of loop-like features are noisy, but still show distinct structures. For example, the first snapshot in Fig.~\ref{Fig.loop.snap} shows some short-lived individual compact brightenings and the last snapshot some elongated structure (at low contrast) from the bottom left to the top right. This presence of spatial structures is pointedly different from the snapshots of the diffuse region (top row of Fig.~\ref{FigA2}). The levels of the count rates are comparable in the diffuse region and the loop-like features, hence noise cannot explain the absence of structures in the diffuse region.

From this discussion we conclude that the region with loop-like features (box iii in Fig.~\ref{Fig1_overview}) hosts small-scale distinct features that cannot be identified in the diffuse region, despite the count rates being comparable. We can speculate if this region with loop-like features is in a transition phase from diffuse to more active, like the bright point or vice versa. To draw solid conclusions on this a more comprehensive study of  loop-like features would be required, in particular including data on the magnetic field configuration in the photosphere below.

\end{appendix}
\end{document}